\documentclass[english]{emulateapj}
\usepackage{amsmath}
\usepackage{color}

\makeatletter

\providecommand{\tabularnewline}{\\}

\@ifundefined{textcolor}{}
{%
 \definecolor{BLACK}{gray}{0}
 \definecolor{WHITE}{gray}{1}
 \definecolor{RED}{rgb}{1,0,0}
 \definecolor{GREEN}{rgb}{0,1,0}
 \definecolor{BLUE}{rgb}{0,0,1}
 \definecolor{CYAN}{cmyk}{1,0,0,0}
 \definecolor{MAGENTA}{cmyk}{0,1,0,0}
 \definecolor{YELLOW}{cmyk}{0,0,1,0}
 }

\makeatother

\usepackage[normalem]{ulem}  

\usepackage{babel}

\usepackage{natbib}

\shorttitle{Chandra Observations of Millisecond Pulsars}
\shortauthors{Mahmoodifar and Strohmayer}

\begin{document}

\title{Where Are the r-modes? {\it Chandra} Observations of
  Millisecond Pulsars}

\author{Simin Mahmoodifar and Tod Strohmayer\\ {\normalfont 
Astrophysics Science Division and Joint Space-Science Institute, NASA's 
Goddard Space Flight Center, Greenbelt, MD 20771, USA} } 

\begin{abstract}

We present the results of {\it Chandra} observations of two
non-accreting millisecond pulsars, PSRs J1640$+$2224 (J1640) and
J1709$+$2313 (J1709), with low inferred magnetic fields and spin-down
rates in order to constrain their surface temperatures, obtain limits
on the amplitude of unstable $r$-modes in them, and make comparisons
with similar limits obtained for a sample of accreting low-mass X-ray binary (LMXB) neutron
stars. We detect both pulsars in the X-ray band for the first
time. They are faint, with inferred soft X-ray fluxes ($0.3-3$ keV) of
$\approx$ $6\times10^{-15}$ and $3\times 10^{-15}$ erg
cm$^{-2}$ s$^{-1}$ for J1640 and J1709, respectively.  Spectral
analysis assuming hydrogen atmosphere emission gives global effective
temperature upper limits ($90\%$ confidence) of $3.3 - 4.3 \times
10^5$ K for J1640 and $3.6 - 4.7 \times 10^5$ K for J1709, where the
low end of the range corresponds to canonical neutron stars ($M=1.4
M_{\odot}$), and the upper end corresponds to higher-mass stars ($M=2.21
M_{\odot}$). Under the assumption that $r$-mode heating provides the
thermal support, we obtain dimensionless $r$-mode amplitude upper
limits of $3.2 - 4.8 \times 10^{-8}$ and $1.8 - 2.8 \times 10^{-7}$
for J1640 and J1709, respectively, where again the low end of the
range corresponds to lower-mass, canonical neutron stars ($M=1.4
M_{\odot}$). These limits are about an order of magnitude lower than
those we derived previously for a sample of LMXBs, except for the accreting millisecond X-ray pulsar (AMXP) SAX
J1808.4$-$3658, which has a comparable amplitude limit to J1640 and
J1709.

\end{abstract}
\section{Introduction}

Neutron stars (NSs) contain the densest matter in the universe outside
of black holes, making them unique laboratories for the study of cold
ultra-dense matter. Their cores may contain exotic forms of matter
postulated to exist at supranuclear densities, including hyperons,
strange quark matter, color superconducting quarks, and pion or kaon
condensates. While constraints on the masses and radii of NSs are
important for constraining the equation of state (EOS) of ultra-dense
matter, observations of their dynamic properties, such as spin and
thermal evolution, as well as their oscillation modes, are also
important and could potentially be more efficient at discriminating
between different phases of dense matter.

A key process that can affect the spin evolution of fast-rotating
pulsars is the instability associated with the $r$-modes of neutron
stars (Bildsten 1998; Andersson et al. 1999). These modes are axial
pulsations in which the restoring force is the Coriolis force
resulting from rotation of the star. The $r$-modes are unstable due to
the emission of gravitational radiation. Since they are retrograde in
the corotating frame of the star and appear prograde to a distant
observer, they emit positive angular momentum in the form of
gravitational waves, but since they have negative angular momentum in
the corotating frame of the star, their amplitude grows due to
emission of gravitational radiation. They may play a role in limiting
the spin rates of neutron stars, since the gravitational radiation
they emit carries away angular momentum and rotational kinetic
energy. For example, it remains puzzling that the spin frequency
distribution of Accreting Millisecond X-ray Pulsars (AMXPs) appears to
cut off well below the mass-shedding limit of essentially all
realistic neutron star EOSs (Chakrabarty et al. 2003; Patruno
2010). White \& Zhang (1997) suggested that magnetic braking could be
responsible for halting the spin-up, while Bildsten (1998) proposed
that it might be due to the emission of gravitational radiation, as
could be produced, for example, by an unstable $r$-mode.

\begin{figure*}[btp]
\begin{center}
\includegraphics[width=3.7in,
height=2.7in]{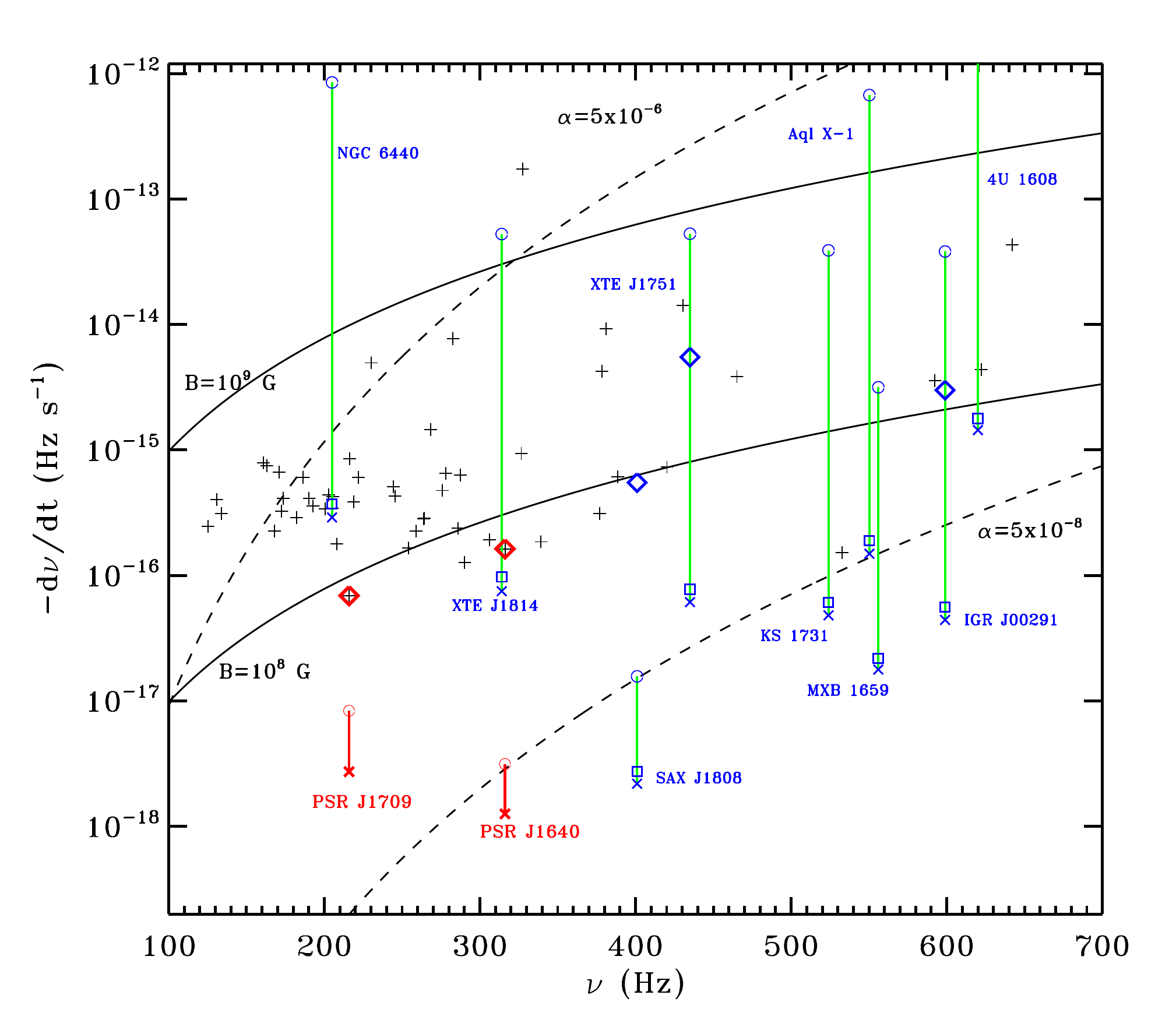}
\caption{\label{fig:nu_ndot} Limits on the spin-down rates due to an
  {\it r}-mode torque for nine LMXB systems (blue symbols and vertical
  green bars, adapted from MS13) and two non-accreting MSPs (red
  symbols and vertical bars, this work) are shown in the $\dot{\nu}$
  vs. $\nu$ plane. The $\dot\nu$ limits for LMXB systems are shown for
  NS masses of $1.4$ ($\times$ symbols), $2.0$ (square symbols), and
  $2.21\, M_{\odot}$ (circle symbols). For LMXB systems with measured,
  quiescent spin-down rates, the values are marked with the blue
  diamonds. The $r$-mode spin-down limits derived for PSRs J1640 and
  J1709 for $1.4$ ($\times$ symbols) and $2.21\, M_{\odot}$ (circle
  symbols) are shown in red. The observed spin-down rates for these
  pulsars are given by the red diamond symbols. For additional
  context, MSPs from the ATNF pulsar database are marked with the
  black $+$ symbols (see MS13 for further details).}
\end{center}
\end{figure*}

Whether or not the $r$-mode instability has a significant effect on
the spin evolution of neutron stars strongly depends on the maximum
amplitude to which they can grow. If their amplitudes
are high enough they may be detectable by the new generation of
gravitational wave detectors, such as advanced-LIGO, but even a
small-amplitude $r$-mode can significantly limit the spin rate of a
neutron star. The $r$-modes are unstable as long as their
gravitational radiation timescale, $\tau_G$, is less than their
damping timescale, $\tau_{damp}$. This competition leads to an
instability window in the $\nu$-T plane---where $\nu$ is the NS's spin
frequency and T is its core temperature---given by a critical curve
that is defined by $\tau_G=\tau_{damp}$. Interestingly, the
instability window is sensitive to the phase of dense matter
comprising the NS (Alford et al. 2012a).

In addition to braking the stars' rotation the $r$-modes also
influence their thermal evolution by depositing heat into the star via
the dissipation of pulsation energy. Both of these processes depend
strongly on the $r$-mode amplitude.  Several mechanisms to dampen the
$r$-modes have been examined, including a viscous boundary layer at
the core-crust interface (Bildsten \& Ushomirsky 2000; Levin \&
Ushomirsky 2001), bulk viscosity enhanced by a hyperon-rich core
(Haskell \& Andersson 2010), supra-thermal bulk viscosity (Alford et
al. 2010, 2012b), mutual friction due to superfluid vortices (Haskell
et al. 2009, 2014), mode coupling (Arras et al. 2003; Bondarescu
et al. 2007, 2009), and magnetohydrodynamic effects (Rezzolla et al.
2000, 2001), but none of these has been shown to eliminate the
$r$-mode instability and they leave the majority of fast-rotating
NSs---such as the class of low-mass X-ray binaries (LMXBs)---inside the
instability window for ``standard'' hadronic matter (Ho et al. 2011;
Haskell et al. 2012; Alford \& Schwenzer 2013; Mahmoodifar \& Strohmayer 2013, hereafter
MS13). The amplitude at which the $r$-mode growth will saturate, and
thereby how much it can affect the spin and thermal evolution of NSs,
is still unknown, but it is crucial for understanding the presence of
fast-rotating NSs (such as the LMXBs) within the instability window.
Thus, temperature and spin evolution measurements of these NSs can
provide a unique probe of their $r$-mode amplitudes and physics, as
well as its potential for constraining the interior properties of NSs.

In MS13 we determined upper limits on the amplitude of $r$-modes, and
their gravitational-radiation-induced spin-down rates, in LMXB NSs
under the assumption that the quiescent NS luminosity is powered by
dissipation from a steady-state $r$-mode. We showed that upper limits
on dimensionless $r$-mode amplitudes in LMXB NSs are in the range of
$10^{-8}$ to $10^{-6}$ for lower mass (1.4 and 2.0 $M_{\odot}$) NS
models and they are larger for high mass (2.21 $M_{\odot}$) models
which support fast (direct Urca) neutrino emission in their cores. For
the three AMXPs, a subset of the LMXBs, with known quiescent spin-down
rates (marked with the blue diamond symbols in Fig. 1) these limits
suggest that $\sim1\%$ of the observed rate can be due to an
$r$-mode. While there are currently only three robust spin-down
measurements for AMXPs (Patruno \& Watts 2012), spin-down rates are
known for a much larger sample of millisecond pulsars (MSPs; shown by the
black $+$ symbols in Fig. 1), and some of these rates are comparable
to the estimates of the $r$-mode spin-down rates in low-mass NS
models.  This suggests the interesting possibility that gravitational
radiation due to $r$-modes in these sources might have a significant
contribution to their spin-down. If so, it may be that some of them
have smaller magnetic fields than their values inferred from magnetic
dipole spin-down alone. However, surface temperatures with which to
constrain the $r$-mode amplitudes are presently not known for many of
these pulsars. To date, only three MSPs have a
global surface temperature estimate. These are the nearby MSPs, PSR
J0437$-$4715 (Durant et al. 2012, hereafter J0437) and PSR
J2124$-$3358 (Rangelov et al. 2017, hereafter J2124), with quite low
surface temperature estimates of $\approx 1 - 4 \times 10^5$ K, and
$0.5 - 2.1\times 10^5$ K for J0437 and J2124, respectively. These
estimates were derived from far UV HST observations. We note that
J0437 is a slow spinner at 173.7 Hz and is likely outside the nominal
instability window. Lastly, PSR J1231$-$1411 (Schwenzer et al. 2017,
hereafter J1231) has an upper limit on the surface temperature of
$1.7\times 10^5$ K, that was obtained using {\it Chandra, XMM-Newton}
and {\it SUZAKU} observations. Based on the discussion above we
proposed to observe with {\it Chandra} a small number of MSPs with low
spin-down rates but faster spin frequencies than J0437, in order to
constrain their core temperatures and the effect of $r$-modes on their
spin evolution. We obtained {\it Chandra} observations for two
targets, PSRs J1640$+$2224 (hereafter, J1640) and J1709$+$2313
(hereafter, J1709), detecting both for the first time in the X-ray
band. Both objects are long-period (see Table 1) neutron star binaries
hosting low-mass white dwarf companions. Age estimates from white
dwarf cooling indicate that the systems are very old, 5 - 7 Gyr
(Desvignes et al. 2016; L{\"o}hmer et al. 2005; Lundgren et
al. 1996). Their ages, combined with the long orbital periods, argues
that these objects have likely not seen any significant accretion,
that is, mass transfer due to Roche lobe overflow, in $\approx 100$
Myr or longer (Willems \& Kolb 2002). Such systems are therefore good
laboratories for exploring NS heating processes, such as $r$-mode
reheating.  This paper is organized as follows. In section 2 we
summarize the observations and data analysis. In section 3 we estimate
the $r$-mode amplitude and spin-down rates in these sources. Finally,
we provide some conclusions in section 4.

 \section{Chandra Observations and Data analysis \label{sec:chandra_obs}} 

\begin{figure*}[btp]
\begin{center}
\begin{tabular}{lr}
\includegraphics[width=0.5\textwidth]{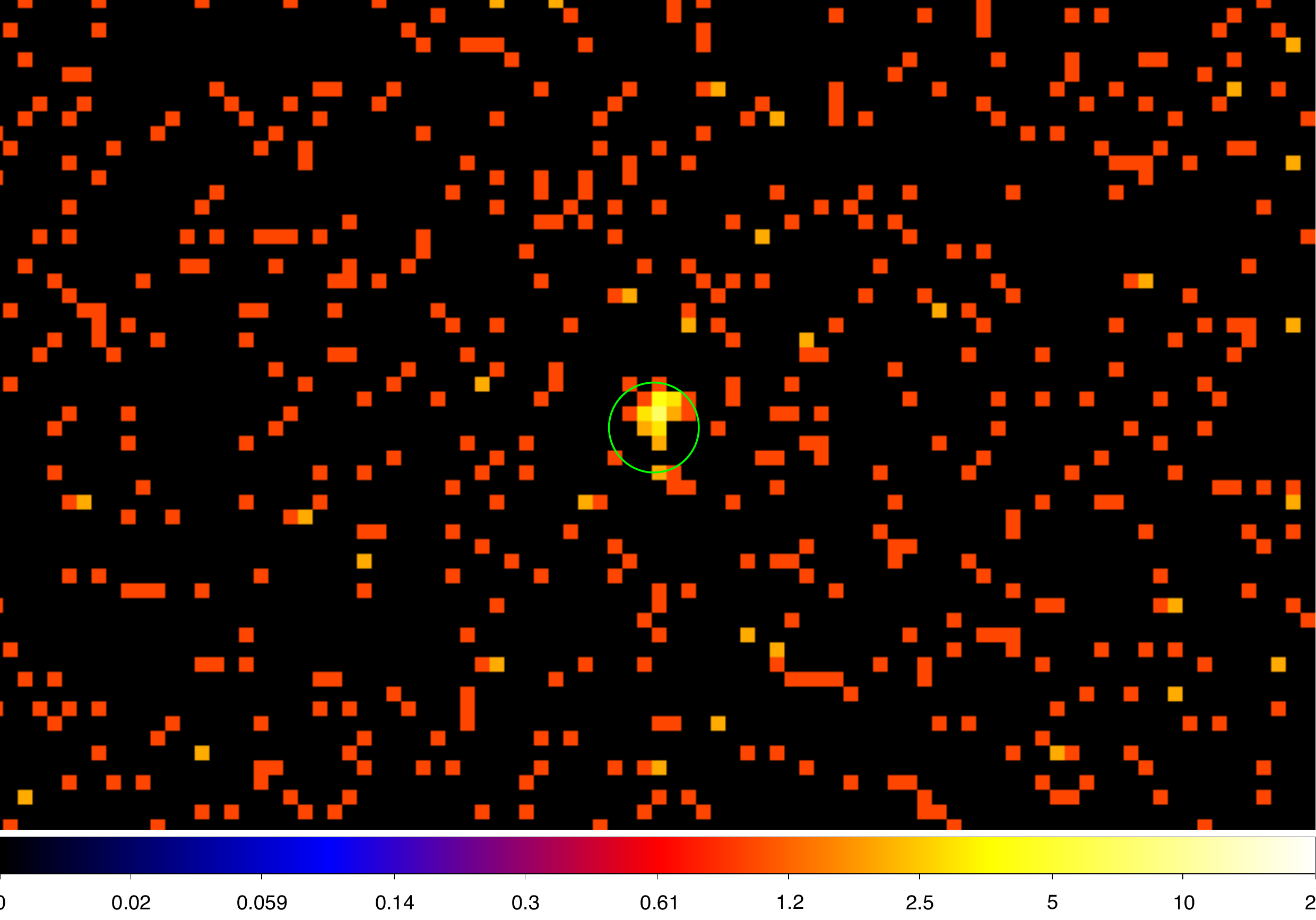}&
\includegraphics[width=0.5\textwidth]{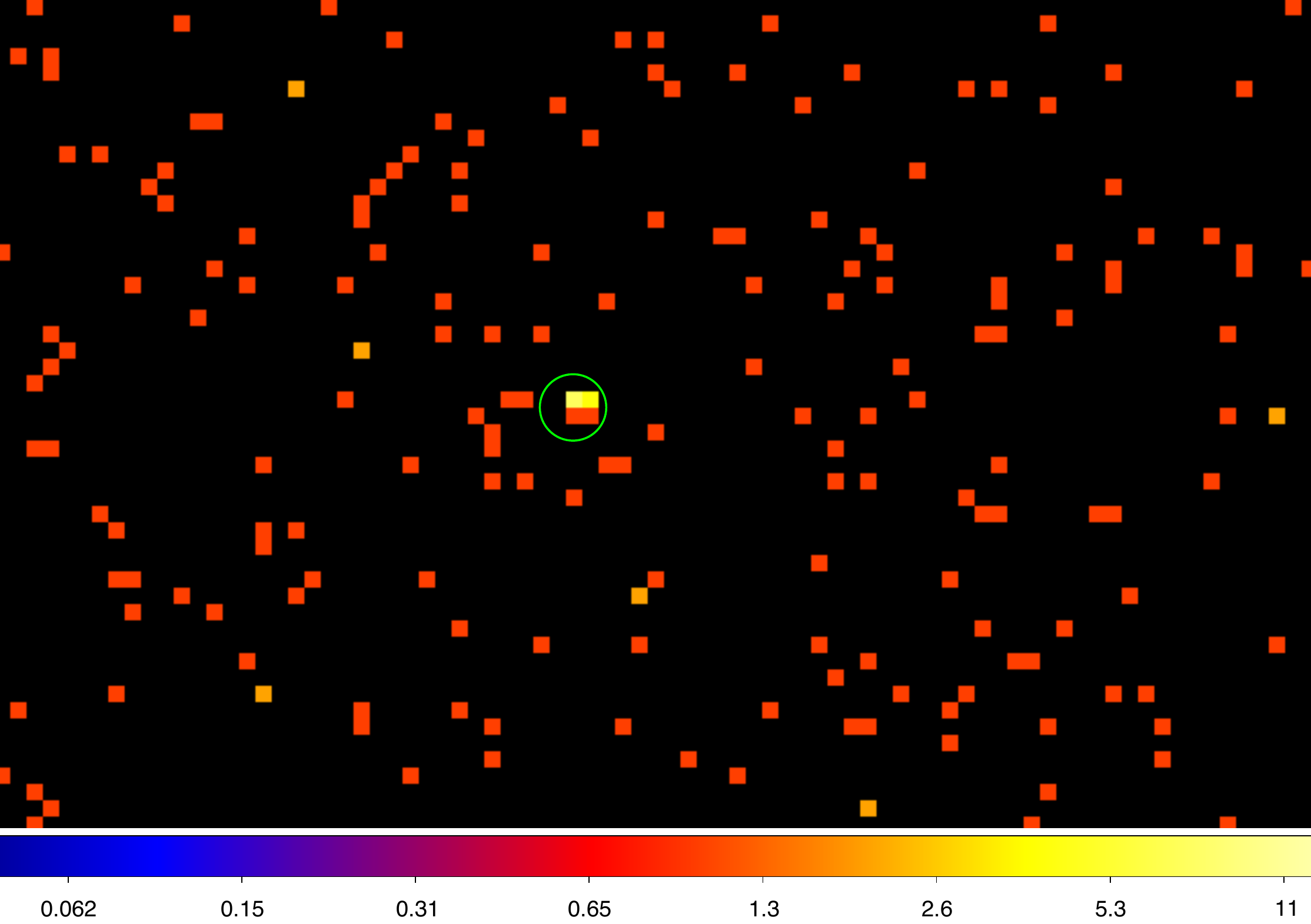} \\
\end{tabular}
\caption{Left: The image of the ACIS-S exposure ($\sim$40794 s) around
  PSR J1640+2224. The green circle shows the $1.5''$ region centered
  on the position of the source (RA: $16^h40^m16.7^s$ Dec: $22^{\circ}
  24' 8.9''$ ). Right: merged image of the two ACIS-S exposures
    (29716 s and 26758 s) around PSR J1709+2313. The green circle
    shows the $1.5''$ region centered on the position of the source
    (RA: $17^h9^m5.8^s$ Dec: $23^{\circ} 13' 27.8''$)}
\label{fig:ACIS_image}
\end{center}
\end{figure*}

\begin{table*}
\begin{center}
\caption{Chandra observations of PSR J1640$+$2224 and PSR J1709$+$2313}
\scalebox{1}{
\begin{tabular}{ccccccccc}
\hline
\hline
 &Obs-ID&Date&Exposure&$\nu_{spin}$& $\frac{d\nu}{dt}|_{obs}$ & $P_b$&$d$ &$n_H$ \\
 &&&(s)& (Hz)&(Hz s$^{-1}$)&(days)&(kpc) &($\times 10^{22}$cm$^{-2}$)\\
\hline
 PSR J1640+2224 &16746&2015 Dec 23&40794& 316.12&$-1.63\times 10^{-16}$&175.46&1.50& 0.044  \tabularnewline
 PSR J1709+2313&16747&2016 Mar 12&26758& 215.93 & $-6.9\times 10^{-17}$&22.71&2.18 &0.046  \tabularnewline
 &18801&2016 Mar 10&29716&&&&&\\
 \hline
\end{tabular}
}
\end{center}
\tablecomments{\footnotesize{{\it Chandra} observation ID, date of
    observation, exposure time, spin frequency, observed spin-down
    rate from pulsar timing, orbital period, distance to the source
    and hydrogen column density. The spin-down rates and distance
    values are from the ATNF Pulsar Catalogue (Manchester et al. 2005;
    Yao et al. 2017)}}
\label{table:obs_info}
\end{table*}

J1640 and J1709 were observed with the {\it Chandra X-ray Observatory}
Advanced CCD Imaging Spectrometer (ACIS-S). The observation IDs,
dates, exposures, and parameters relevant for our spectral analysis
for each of these sources are presented in Table 1.  We note that
neither of these sources had a prior detection with either {\it
  Chandra}, {\it XMM-Newton}, {\it Swift} or indeed {\it ROSAT}.

In the following subsections we present our data analysis and methods
for constraining the global surface temperature of these pulsars.  We
note that some MSPs show a soft X-ray thermal emission component that
is very likely associated with small polar caps heated by
pulsar-induced return currents striking the magnetic poles (Zavlin
2006).  A good example of this is PSR J0437 (Bogdanov et al
2007). Where X-ray pulsations are detected they have been associated
with this spectral component as well (Bogdanov 2013).  The inferred
temperature of this component, $\sim 200$ eV, tends to be higher than
the global surface temperature (Durant et al. 2012), but the emission
arises from only a small fraction of the NS surface, consistent with
the polar cap interpretation. As we describe below, it is likely that
our targets may have such polar cap emission, but it is important to
recognize that global temperature constraints can still be derived.
These stars have very low magnetic fields $\le 10^{8}$ G that do not
appreciably distort or beam the atmospheric emission that must be
present over their entire surfaces.

\begin{table*}
\begin{center}
\caption{Chandra ACIS-S best spectral models for PSR J1640$+$2224}
\scalebox{1}{
\begin{tabular}{ccccccccc}
\hline
\hline
Model & M & R  & norm1 & Log T$_{eff,1}$ & norm2& Log T$_{eff,2}$&Model predicted rate& Cstat/dof\\
&M$_{\odot}$&(km)&&(K)&&(K)&(cts/s)&\\
\hline
{\it wabs(nsatmos+nsatmos)}& 1.4 & 11.5  & $1.5\times10^{-4}$& 6.32 & 1.0 & 5.41 $\pm$ 0.12 & $7.3\times10^{-4}$ & 101.5/197\\
\\
{\it wabs(nsatmos+nsatmos)}& 2.21 & 10 & $1.0\times10^{-4}$& 6.46 & 1.0 & 5.53 $\pm$ 0.12 & $7.3\times10^{-4}$ & 101.4/197\\
\\
\hline
\end{tabular}
}
\end{center}
\tablecomments{\footnotesize{The values of $N_H$ and the distance to the source are fixed at 0.044$\times10^{22}$ cm$^{-2}$ and 1.5 kpc, respectively.}}
\label{table:J1640_parameters}
\end{table*}


\subsection{PSR J1640+2224}

We selected source counts in a $1.5''$ circular region around the
target coordinates. The background is selected from three source-free
$9''$ radii regions distributed around the source. The data processing
was performed using CIAO 4.8. The left panel of Fig.~\ref{fig:ACIS_image}
shows the image of the ACIS-S exposure around PSR J1640+2224. The
green circle shows the $1.5''$ region centered on the position of the
source (RA: $16^h40^m16.7^s$ Dec: $22^{\circ} 24' 8.9''$ ). In total
there are 32 counts in the $\sim$40.8 ksec exposure (considering all
1024 detector energy channels). This is a very strong detection for
such an exposure, given {\it Chandra's} superb angular resolution and
low background, and represents the first detection of J1640 in the
X-ray band. Although the total number of observed source counts is not
high enough for a detailed spectral analysis, it is still sufficient
to constrain the surface temperature of the star.
 
The spectral analysis is performed using XSPEC version 12.8.2 (Arnaud
1996). All but two counts in the source extraction region have an
energy below 3 keV, which is consistent with the expected, soft X-ray
spectrum of other MSPs. Thus, we consider only the $0.1-3$ keV energy
band (ACIS channels $8-205$) for spectral analysis. The net count rate
in the source region is $7.21\times 10^{-4} \pm 1.34\times 10^{-4}$
cts/s, with a total of 30 counts. To take into account the effects of
interstellar absorption we adopt a galactic absorbing column, $n_{H}$,
determined with HEASARC's ``$n_H$-tool,'' and we use the {\it wabs}
photoelectric absorption model in XSPEC (Morrison \& McCammon 1983).

Since the number of counts in each channel is small, in our spectral
fitting we use the C-statistic ({\it cstat} in XSPEC), which is the
maximum likelihood-based statistic appropriate for Poisson data (Cash
1979), instead of the $\chi^2$ statistic. We note that in the limit of
large numbers of counts this statistic asymptotes to $\chi^2$. The
C-statistic can be used regardless of the number of counts in each
bin, and an approximate goodness-of-fit measure for a given value of
the {\it cstat} statistic can be obtained by dividing the observed
statistic by the number of degrees of freedom, which should be of
the order of 1 for good fits.

To model the spectrum we use the NS hydrogen atmosphere model {\it
  nsatmos} in XSPEC (Heinke et al. 2006). We employ this model rather
than a blackbody, because during the evolution of MSPs they accrete a
substantial amount of matter and it is expected that they have an
atmospheric layer on their surfaces. Fitting a blackbody to a MSP
spectrum results in a smaller inferred emission area and a higher
temperature compared to fitting a hydrogen atmosphere model (Zavlin
2006; Bogdanov 2013). To constrain the global surface temperature of
the star we first fit the spectrum with an absorbed hydrogen
atmosphere model ({\it wabs*nsatmos}), fixing the NS mass, radius, and
$n_H$, and find the best values for the other two parameters, namely
the temperature and the normalization (defined as the fraction of the
NS surface that is emitting, hereafter, norm1). We find that the
inferred temperature and emitting area are consistent with the
existence of a hotspot on the NS surface. However, in order to
constrain the NS's core temperature, we need an upper limit on the
global surface temperature of the star, which is expected to be lower
than the hotspot temperature. Therefore we fix all the hotspot model
parameters to their best-fit values and add another hydrogen
atmosphere model component to account for the global surface emission,
and fit the spectrum again. The full XSPEC model in this case is {\it
  wabs*(nsatmos+nsatmos)}. For the second, surface emission {\it
  nsatmos} model, we fix the values of $n_H$, NS mass, and radius to
the same values as for the hotspot component, and we also fix the
value of the normalization parameter for this component, norm2, to
1.0-norm1, and in this case ${\rm norm2} \sim 1$. Then we fit the
spectrum again to constrain the value of the second (surface)
temperature parameter.

\begin{figure*}[btp]
\begin{center}
\begin{tabular}{lr}
\includegraphics[width=0.5\textwidth]{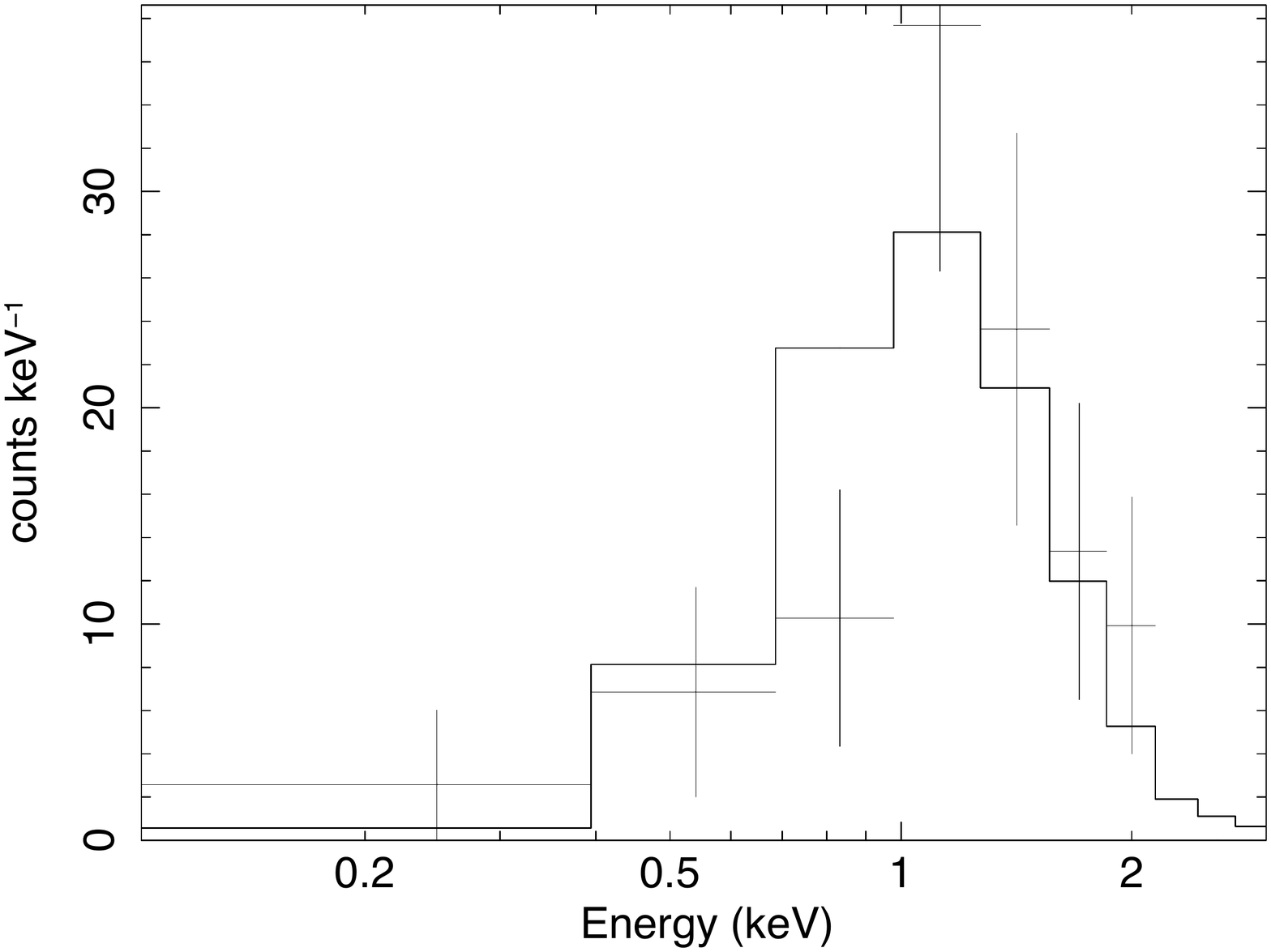}&
\includegraphics[width=0.5\textwidth]{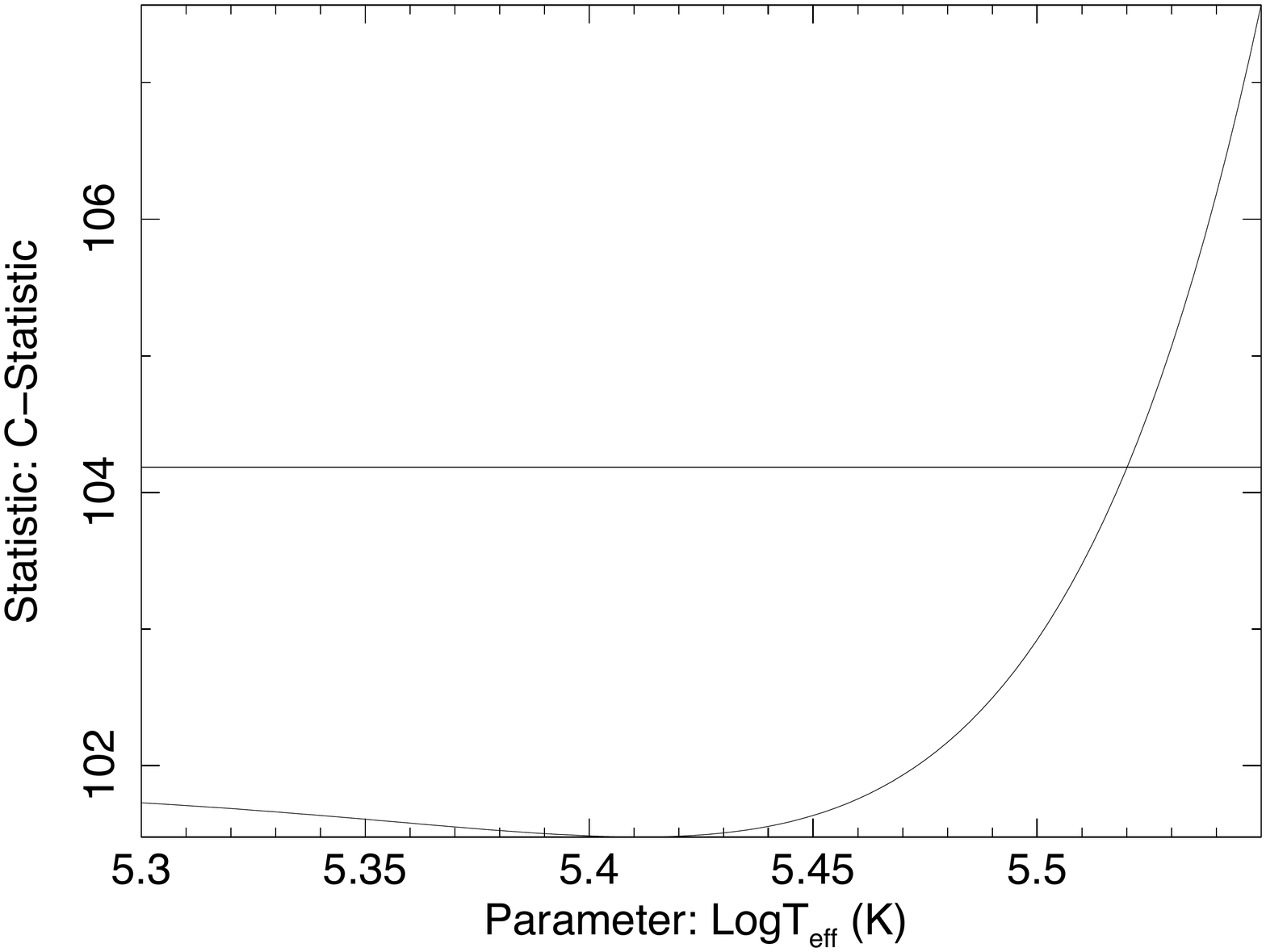} \\
\end{tabular}
\caption{Left: Chandra ACIS-S X-ray spectrum and the folded model
  ({\it wabs*(nsatmos+nsatmos})) for J1640 in the 0.1-3 keV energy
  band. The NS mass and radius in these models are fixed at 1.4
  M$_{\odot}$ and 11.5 km, respectively. Right: confidence region for
  the second temperature parameter in the fit of the two absorbed hydrogen
  atmosphere models. The horizontal line shows the 90\% confidence
  region.}
\label{fig:J1640_count_steppar1.4}
\end{center}
\end{figure*}

To put an upper limit on the global surface temperature of the star we
fix all the model parameters except the second temperature and vary
that parameter using the {\it steppar} command in XSPEC until the
C-statistic changes by 2.706, equivalent to the 90\% confidence region
for a single interesting parameter. The result is shown in the right
panel of Fig.~\ref{fig:J1640_count_steppar1.4}.  We do the same
procedure for two different NS models with masses and radii of 1.4
M$_{\odot}$ and 11.5 km, and 2.21 M$_{\odot}$ and 10 km\footnote
{Here, radii are computed using the Akmal-Pandharipande-Ravenhall
  (APR) equation of state (Akmal et al. 1998).}. The resulting
spectral model parameters for PSR J1640+2224 are given in Table 2.

We tried three different hydrogen atmosphere models ({\it nsa}, {\it nsagrav}, and {\it nsatmos}), and a helium atmosphere model ({\it nsx}) in XSPEC in our spectral analysis, and for the range of parameters that we are interested they give very similar results. We have chosen to use the {\it nsatmos} model since it has a separate parameter (``norm," described above) for the emitting area that makes it easier to work with. With the He atmosphere model, fitting only one component (i.e. {\it wabs*nsx}) gives a comparable fit to the one-component H atmosphere model ({\it wabs*nsatmos}), but after adding the second component to the fit to constrain the global surface temperature, it does not give as good a fit as the two {\it nsatmos} model fits in some cases, since the temperature of the second component in {\it nsx} cannot go below Log $T_{eff}=5.5$. 

We also tried fitting both {\it nsatmos} models at the same time (still keeping the second normalization parameter fixed at 1) and it gave similar results as fitting two components separately. The differences in the values of the second temperature parameter (global surface temperature) are very small and within the error bar (for example, the difference is less than 1 eV in the case of J1640).

\subsection{PSR J1709+2313}

\begin{figure*}[btp]
\begin{center}
\begin{tabular}{lr}
\includegraphics[width=0.5\textwidth]{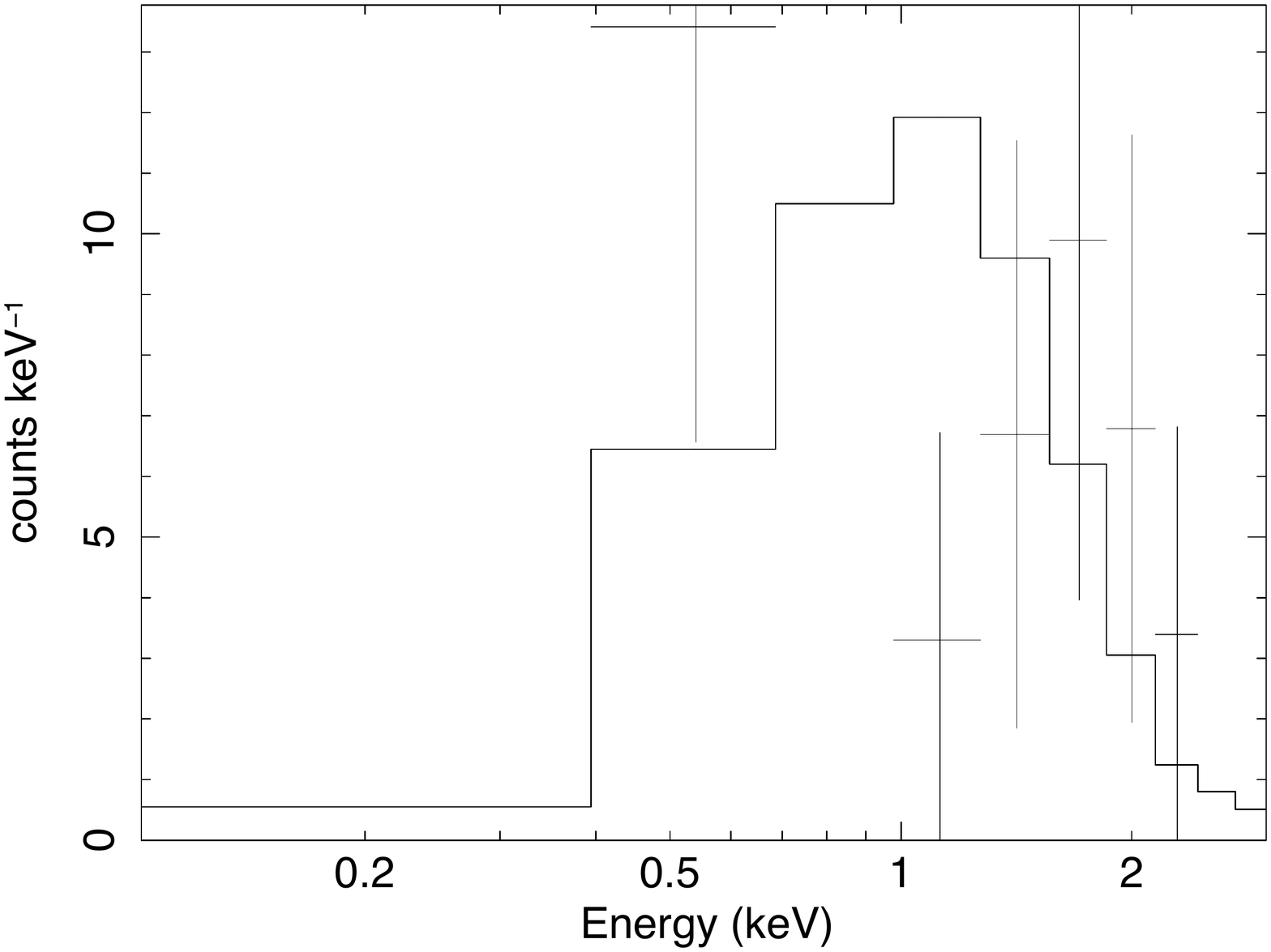}&
\includegraphics[width=0.5\textwidth]{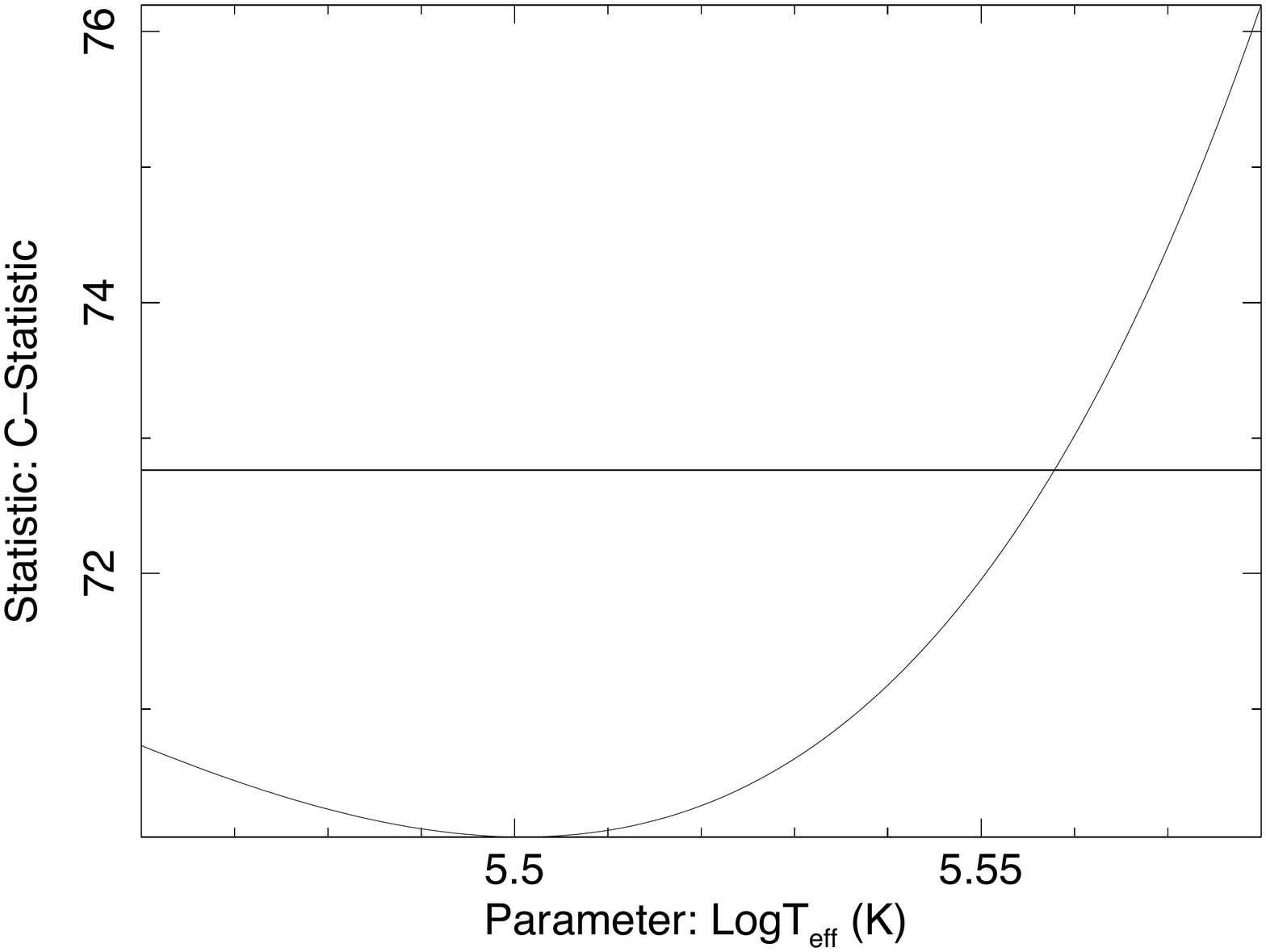} \\
\end{tabular}
\caption{Left: Chandra ACIS-S X-ray spectrum and the folded model
  ({\it wabs*(nsatmos+nsatmos})) for J1709 in the 0.1-3 keV energy
  band. The NS mass and radius in these models are fixed at 1.4
  M$_{\odot}$ and 11.5 km, respectively. Right: Confidence region for
  the second temperature parameter in the fit of the two absorbed hydrogen
  atmosphere models. The horizontal line shows the 90\% confidence
  region.}
\label{fig:J1709_count_steppar1.4}
\end{center}
\end{figure*}

The observation of J1709 was done in two parts, for a total exposure
of 56.5 ksec (see Table 1 for details).  Similarly to J1640 we
selected the source counts in a $1.5''$ circular region around the
target coordinates and the background from 3 source-free $9''$ radii
regions distributed around the source. The green circle in the right
panel of Fig.~\ref{fig:ACIS_image} shows the $1.5''$ region centered
on the position of the source (RA: $17^h9^m5.8^s$ Dec: $23^{\circ} 13'
27.8''$ ). After extracting spectra and responses for each observation
we used {\it combine-spectra} in CIAO 4.8 to coadd the spectra and
responses.  In total there are 13 counts in the 0.1-3 keV band, and
the net count rate is $2.19\times 10^{-4} \pm 6.39\times 10^{-5}$
cts/s during 56.5 ksec total exposure. There are a few photons above 9
keV that have been ignored in our spectral analysis. We carried out a
similar procedure as for J1640 to fit the spectrum with an absorbed
hydrogen atmosphere model, {\it wabs*(nsatmos+nsatmos)} in XSPEC. The
results of our best fits for two NS models with a canonical and a high-mass are given in Table 3. The left panel of
Fig.~\ref{fig:J1709_count_steppar1.4} shows the data and folded model
for a 1.4 M$_{\odot}$ star, and the right panel shows the upper limit
on the global surface temperature, similar to
Fig.~\ref{fig:J1640_count_steppar1.4}.

\begin{table*}
\begin{center}
\caption{Chandra ACIS-S best spectral models for PSR J1709+2313}
\scalebox{1}{
\begin{tabular}{ccccccccc}
\hline
\hline
Model & M & R & norm1 & Log T$_{eff,1}$ & norm2& Log T$_{eff,2}$&Model Predicted Rate& Cstat/dof\\
&M$_{\odot}$&(km)&&(K)&&(K)&(cts/s)&\\
\hline
{\it wabs(nsatmos+nsatmos)}& 1.4 & 11.5 & $3.9\times10^{-5}$ & 6.41 & 1.0 & 5.50 $\pm$ 0.04 & $2.63\times10^{-4}$ & 70.1/197\\
\\
{\it wabs(nsatmos+nsatmos)}& 2.21 & 10 & $4.2\times10^{-5}$ & 6.50 & 1.0 & 5.61$\pm$ 0.04 & $2.57\times10^{-4}$ & 70.6/197\\
\\
\hline
\end{tabular}
}
\end{center} \tablecomments{\footnotesize{The values of $n_H$
and the distance to the source are fixed at 0.046$\times10^{22}$
cm$^{-2}$ and 2.18 kpc, respectively.}}
\label{table:J1709_parameters}
\end{table*}

\section{Constraining the $r$-mode amplitude and spin-down 
rate\label{sec:r-mode}}

In order to constrain the amplitude of the $r$-mode oscillations in
these pulsars, we first need to compute the core temperatures using
the surface temperature limits we obtained from spectral fitting in
section 2. We use the following equation that
relates the effective surface temperature of the star, $T_{eff}$, to
the internal temperature, $T_b$, which is the temperature at a
fiducial boundary at $\rho_b=10^{10}$ g cm$^{-3}$ for a fully accreted
envelope and is valid for $T_b \le 10^8$ K (Potekhin et al. 1997),
\begin{equation}
(\frac{T_{eff}}{10^6 K})^4=(\frac{g}{10^{14} cm s
    ^{-2}})(18.1\frac{T_b}{10^9 K})^{2.42}\label{eq:T_b potekhin}
\end{equation}
where $g=GM/(R^2 \sqrt{1-r_g/R})$ is the surface gravity and
$r_g=2GM/c^2$.  Here we assume that the neutron star's core is
isothermal and since the thermal conductivity of the crust is high
(Brown \& Cumming 2009) we have $T_{core}=T_b$ to good approximation.  For
a partially accreted envelope the core temperature would be slightly
higher, but the difference is negligible for our cases (see equation
A$9$ in Potekhin et al. (1997)). The values of the core temperatures
are given in Table 5.  Having the core and surface
temperatures we can now determine the photon and neutrino luminosities
using the following equations (see MS13):

\begin{equation}
L_{\gamma}=4\pi R^2 \sigma T_{eff}^4
\end{equation}

where $R$ and $T_{eff}$ are the stellar radius and surface
temperature, respectively, and

\begin{equation}
L_{\nu}=\frac{4\pi R_{DU}^3 \Lambda_{QCD}^3
  \tilde{L}_{DU}}{\Lambda_{EW}^4}T^6+\frac{4\pi R^3 \Lambda_{QCD}
  \tilde{L}_{MU}}{\Lambda_{EW}^4}T^8
\end{equation}

where $T$ is the core temperature, $R_{DU}$ is the radius of the core
where direct Urca neutrino emission is allowed, and $\tilde{L}$ is a
dimensionless parameter given in
Table 4. $\Lambda_{QCD}$ and $\Lambda_{EW}$ are
characteristic strong and electroweak scales with values of 1 GeV and
100 GeV, respectively, that we have used in our calculations.

Assuming that $r$-modes are excited in these neutron stars, we can use
the thermal equilibrium argument discussed in Brown \& Ushomirsky
(2000; MS13) to put an upper limit on their $r$-mode amplitudes. We
note that based on their inferred core temperatures, these pulsars
lie in the uncertainty region of the r-mode instability window (see
Fig. 6 of Schwenzer et al. 2017) and $r$-mode oscillations might be
completely damped in these sources. However, we still compute the upper
limits on their $r$-mode amplitudes assuming that they are
within the unstable region.

We consider that an NS to be in thermal equilibrium
when the reheating due to $r$-mode dissipation is balanced by cooling
due to core neutrino emission and the thermal photon emission
from the surface of the star, $W_d=L_{\nu}+L_{\gamma}$. The reheating
due to $r$-mode dissipation can be computed using the equation $W_d=-2
E_c/\tau_{GR}$ (MS13).

\begin{table*}
\begin{center}
\caption{Parameters of the Neutron Star Models\label{tab:viscosity-parameters}}
\scalebox{1}{
\begin{tabular}{ccccccc}
\hline
\hline
Neutron Star & Shell & $R (km)$&$\Omega_K (Hz)$&$\tilde{I}$& $\tilde{J}$ & $\tilde{L}$\\
\hline
NS $1.4\, M_{\odot}$ & Core & $11.5$&$6020$&$0.283$ & $1.81\times10^{-2}$ &  $1.91\times10^{-2}$ \\
NS $2.21\, M_{\odot}$ & m.U. core &$10.0$&$9310$&$0.295$& $2.02\times10^{-2}$ & $1.29\times10^{-2}$\\
 & d.U. core &5.9& & & &  $2.31\times10^{-5}$ \\
\hline
\end{tabular}}
\end{center}
\tablecomments{\footnotesize{Radius, Kepler frequency and
    radial integral parameters that appear in the moment of inertia,
    angular momentum of the mode, and neutrino luminosity for
    different neutron star models considered in this work (Alford et
    al . 2010, 2012a). For the $2.21\, M_{\odot}$ model the neutrino
    luminosity parameter is different in the inner core where direct
    Urca processes are allowed, therefore these values are given
    separately in the last row.}}
\label{table:NS_models}
\end{table*}

\begin{table*}
\begin{center}
\caption{NS temperatures and luminosities, and $r$-mode amplitude and
  spin-down rates} \scalebox{1}{
\begin{tabular}{cccccccccc}
\hline
\hline
 &$\nu_{spin}$& T$_{eff}$ & $g/10^{14}$ & T$_b$ & L$_{\gamma}$ & L$_{\nu}$& $\alpha$&$\frac{d\Omega}{dt}|_G$& $\frac{d\Omega}{dt}|_{obs}$\\
&(Hz)&(K)&(cm/s$^2$)&(K)&(erg/s)&(erg/s)&&(Hz/s)&(Hz/s)\\
\hline
J1640+2224\\
\hline 
1.4M$_{\odot}$, 11.5km&316.12&$3.31\times10^{5}$ & 1.755 & $7.05\times10^{6}$&$1.13\times10^{31}$ & $2.15\times10^{22}$ & $3.2\times10^{-8}$ &$-1.3\times10^{-18}$&$-1.63\times10^{-16}$\\
\\
2.21M$_{\odot}$, 10km&&$4.28\times10^{5}$  & 4.976 & $7.01\times10^{6}$&$2.40\times10^{31}$& $9.18\times10^{30}$ & $4.8\times10^{-8}$ &$-3.4\times10^{-18}$&$-1.63\times10^{-16}$\\
\hline
J1709+2313\\
\hline 
1.4M$_{\odot}$, 11.5km&215.93 &$3.61\times10^{5}$ & 1.755 & $8.14\times10^{6}$& $1.61\times10^{31}$& $6.77\times10^{22}$ & $1.8\times10^{-7}$ &$-2.7\times10^{-18}$&$-6.9\times10^{-17}$\\
\\
2.21M$_{\odot}$, 10km&&$4.68\times10^{5}$ & 4.976 & $8.12\times10^{6}$& $3.42\times10^{31}$& $2.21\times10^{31}$ & $2.8\times10^{-7}$ &$-8.4\times10^{-18}$&$-6.9\times10^{-17}$\\
\\
\hline
\end{tabular}
} 
\end{center}
\tablecomments{\footnotesize{The values of T$_{eff}$ are upper limits
    on surface temperatures and therefore are slightly higher than the
    T$_{eff,2}$ given in Tables 2 and 3; $g$ is the surface
      gravity, $T_b$ is the core temperature, $L_{\gamma}$ and
      $L_{\nu}$ are photon and neutrino luminosities, $\alpha$ is the
      upper limit on the saturation amplitude of the r-modes, and
      $\frac{d\Omega}{dt}|_G$ and $\frac{d\Omega}{dt}|_{obs}$ are
      upper limits on the r-mode-induced spin-down rates and the
      observed spin-down rates of these pulsars, respectively.}}
\label{table:T_alpha}
\end{table*}

Since $W_d$ is a function of $r$-mode amplitude, $\alpha$, one can
place an upper limit on $\alpha$ assuming that $r$-mode dissipation is
the only source of heat inside the star

\begin{equation}
\alpha=\frac{5\times 3^4}{2^8 \tilde{J} M R^3
  \Omega^4}(\frac{L_{\gamma}+L_{\nu}}{2 \pi G})^{1/2}
\end{equation}

Upper limits on $\alpha$ for two different NS models with canonical
and high mass are given in Table 5. These bounds are
about an order of magnitude lower than the bounds obtained for LMXB
sources with comparable frequencies (MS13), because of the lower
temperature limits in these pulsars compared to LMXBs.

To obtain limits on $r$-mode spin-down rates we note that since these
neutron stars are old, if they are inside the $r$-mode instability
window, their $r$-mode amplitudes should be saturated by some
mechanism, meaning $\frac{d\alpha}{dt}=0$, otherwise they would have
been spun down long before. This implies that
$\frac{1}{\tau_V}=-\frac{1}{\tau_G}\frac{1}{1-\alpha^2Q}$ in the
saturation regime, where $Q=3\tilde{J}/2\tilde{I}$ (MS13).  The values
of $\tilde{J}$ and $\tilde{I}$ are given in
Table 4. Therefore the equation for the evolution
of NS spin frequency can be written as (MS13):

\begin{equation} 
\frac{d\Omega}{dt}=2Q\frac{\Omega\alpha^2}{\tau_G(1-\alpha^2Q)}
\end{equation}
 
where

\begin{equation}
\frac{1}{\tau_G}=-\frac{P_G}{2E_c}
\end{equation}

\begin{equation}
E_c=\frac{1}{2}\alpha^2\Omega^2\tilde{J}MR^2
\end{equation}

\begin{equation}
P_G=\frac{2^{17}\times\pi}{5^2\times3^8}\tilde{J}^2GM^2R^6 \alpha^2 \Omega^8
\end{equation}

$E_c$ is the canonical energy of the $r$-mode and $P_G$ is the power
radiated by gravitational waves. Note that $P_G$ here is given for the
$l=m=2$ $r$-mode (MS13) which is the most unstable one. The upper
limits on the $r$-mode-induced spin-down rates, along with the observed
spin-down rates from pulsar timing for these pulsars, are given in
Table 5 and shown in Figure 1. Comparing these
values we find that for J1640 the $r$-mode spin-down cannot be more
than 2\% of the observed spin-down of the pulsar. For J1709 this
number is higher, at 12\%, but we note that this source might not
even have any excited $r$-mode oscillations in it, since it lies
outside but within the uncertainty region of the $r$-mode instability
window for pure hadronic matter.


\section{Discussion\label{sec:discussion}}

We have used {\it Chandra} ACIS-S observations of J1640 and J1709 in
order to constrain their global surface temperatures and put limits on
their $r$-mode amplitudes and corresponding spin-down rates. These
observations represent the first detections of these sources in the
X-ray band.  Unsurprisingly, the objects are faint, with estimated
fluxes ($0.3 - 3$ keV) of $6\times10^{-15}$ and $3\times
  10^{-15}$ erg cm$^{-2}$ s$^{-1}$ for J1640 and J1709, respectively.

We first fit their {\it Chandra} spectra with an absorbed hydrogen
atmosphere model. The inferred temperatures and emitting areas that we
extracted are consistent with the existence of a hotspot on the NS
surface. To obtain a limit on the global surface temperature of the
star, which is much lower than the hotspot temperature, we add
another hydrogen atmosphere model to our fit and fix all the
parameters to the best-fit values we obtained initially, except the
temperature of the second model. We found upper limits on the global
surface temperature of these pulsars that are $\sim 3.3\times
10^5-4.7\times 10^5$K, depending on the assumed NS masses and
radii. Our results are broadly consistent with the results of Gusakov
et al. (2016) where they have given an upper limit of $6\times 10^5$ K
for the redshifted surface temperature of MSPs. Note
that the values of temperatures given here are the temperatures at the
surface of the star, $T_s$, and not the redshifted ones, $T_{\infty}$,
where $T_s=(1+z) T_{\infty}$.

The characteristic spin-down ages of these sources are about
15-16 Gyr (Lorimer 2008), and white dwarf cooling age estimates
  also indicate that they are at least several gigayears old (Desvignes et
  al. 2016).  In all standard cooling models neutron stars cool to
  surface temperatures less than $10^4$ K in less than $10^7$ yr (Page
  2009; Yakovlev \& Pethick 2004). While we have derived {\it upper
    limits} on the surface temperatures for J1640 and J1709, they
  appear to be consistent with the values {\it measured} for J0437 and
  J2124. Taken together these results suggest that the surface
  temperatures of at least some MSPs are significantly higher, given
  their ages, than standard cooling models would suggest.

These high temperatures cannot be attributed to accretion, since the
time it takes for the star to cool to such low temperatures (a few
million years) is much less than the time since accretion has ceased
in the binary system ($> 100$ Myr) (Willems \& Kolb 2002; Kargaltsev
et al. 2004). Therefore, some other source of reheating is likely
operating in these pulsars. Kargaltsev et al. (2004) and Gonzalez \&
Reisenegger (2010) have discussed some internal and external sources
of heat, which includes: (1) dissipation of the energy of differential
rotation caused by frictional interaction between the faster-rotating
superfluid core and the slower-rotating solid crust (Shibazaki \& Lamb
1989; Larson \& Link 1999); (2) energy release due to the cracking of
the crust (Cheng et al. 1992) as well as changes in the rate of
nuclear reactions due to the readjustment of the NS structure to a new
equilibrium state as the star spins down (rotochemical heating;
Reisenegger 1995); (3) the energy release due to the pinning and
unpinning of the vortex lines with respect to the nuclei of the
crystal lattice as the star spins down (vortex creep; Alpar et
al. 1984); and (4) magnetospheric heating (Harding \& Muslimov 2002). In the
case of pulsars that are hot enough to be inside the $r$-mode
instability window, e.g. J1640, then $r$-mode dissipation can provide
another source of reheating.

We used the upper limits on the surface temperature of these pulsars
to first calculate their core temperatures, and then the upper limits
on their $r$-mode amplitudes and their $r$-mode-induced spin-down
rates. The limits on the amplitudes are $\sim (3.2-4.8)\times 10^{-8}$
and $\sim (1.7-2.8)\times 10^{-7}$ for J1640 and J1709, respectively,
depending on the assumed NS mass and radius and the existence of fast
neutrino cooling in the core. These values are about an order of
magnitude lower than the upper bounds obtained for most of the LMXB
sources with comparable frequencies (MS13), because of the lower
temperatures in these pulsars.  Interestingly, the limits for J1640
and J1709 are in line with the upper limits derived for the AMXP SAX
J1808.4$-$3658, which unlike J1640 and J1709, has had recent
accretion. The upper bounds on the $r$-mode spin-down rate for J1640
is $\sim (1.3-3.4)\times 10^{-18}$ and for J1709 is $\sim
(2.7-8.3)\times 10^{-18}$. Comparing these values we find that for
J1640 the $r$-mode spin-down cannot be more than 2\% of the observed
spin-down of the pulsar, and for J1709 it cannot be more than 12\%. As noted previously, both of these systems
  reside near the boundary of the $r$-mode instability window and within
  the uncertainty region, therefore both of them might be $r$-mode
  stable, in which case no particularly new physics, either for
  damping or saturation of the $r$-mode, would be required to explain
  such low upper limits.

Here we assumed that these systems are inside the $r$-mode
  instability window but their $r$-modes are saturated at some small
  amplitudes and are no longer growing. As the star spins down and
  cools, eventually it will leave the unstable region, but if the
  saturation amplitude is very small, as is the case for these
  systems, it can live inside the instability window for a very long
  time (see Fig. 1 and 2 in Alford et al. 2012c). It doesn't seem
  unlikely to find two systems in this phase of their evolution,
  because based on their observed spin-down rates ($1.63 \times
  10^{-16}$ Hz/s for J1640 and $6.9\times 10^{-17}$ Hz/s for J1709) it
  would take billions of years to spin down these pulsars by a few Hertz,
  therefore accretion seems to be the main mechanism that places these
  systems in their current location in or near the instability window.
  Given the limits on the temperatures and $r$-mode saturation
  amplitudes in these systems, the $r$-mode-induced spin-down rate can
  be only a few percent of the observed spin-down rate and it will not
  have a big effect on their evolution.

It remains an open question whether $r$-modes are excited with these
tiny amplitudes in MSPs or if they are completely damped. Although such
low-amplitude $r$-modes will not have much of an effect on the spin evolution
of old pulsars, their dissipation can still provide an extra source of
heat for the NS. Then the question would be which mechanism would be
able to dampen the $r$-modes to such low amplitudes. There are several
mechanisms that have been proposed (see the introduction for a brief
summary), but of these none have been shown to eliminate the $r$-mode
instability or saturate $r$-mode amplitudes at such low values. One
process that could perhaps saturate $r$-mode amplitudes at very low
levels is phase conversion in a multicomponent compact star (Alford et
al. 2015), which can saturate $r$-modes at amplitudes smaller than
$10^{-10}$; however, that also depends on the microscopic and
astrophysical parameters of the star, such as the mass of the quark
core, which should not be too small. Dissipation of $r$-modes at such
low amplitudes in hybrid stars will not provide enough heating to affect
the temperature of the star. Another interesting proposal for
  suppression of the $r$-mode instability at certain stellar
  temperatures is resonant mode coupling and enhanced damping of
  $r$-modes in NSs with a superfluid core (Gusakov et al. 2014; Kantor
  et al. 2016).

Looking ahead, measurements or upper bounds on the surface temperature
of MSPs, especially those with higher spin frequencies, which are
expected to be inside the nominal instability window for hadronic
matter, will provide a valuable tool to study $r$-mode physics and
constrain the properties of neutron star interiors. MSPs with
$T_{eff}^{\infty} > 6\times 10^5$ K would provide strong evidence for
the existence of unstable $r$-modes in such pulsars (Gusakov et
al. 2016), since $r$-mode dissipation would be the most natural source
of heating in the absence of accretion.  \\

Support for this work was provided by the National Aeronautics and
Space Administration through Chandra Award Number GO5-16049Z issued by
the Chandra X-ray Observatory Center, which is operated by the
Smithsonian Astrophysical Observatory for and on behalf of the
National Aeronautics Space Administration under contract
NAS8-03060. S.M.'s research was supported by an appointment to the NASA
Postdoctoral Program at the GSFC.

\vskip 5pt
\section*{References}

{\footnotesize 
\scriptsize

\noindent\hangindent=0.5cm\hangafter=1 Akmal, A., Pandharipande, V.~R., \& Ravenhall, D.~G.\ 1998, \prc, 58, 1804

\noindent\hangindent=0.5cm\hangafter=1 Alford, M.~G., Han, S., \& Schwenzer, K.\ 2015, \prc, 91, 055804

\noindent\hangindent=0.5cm\hangafter=1 Alford, M.~G., Mahmoodifar, S., \& Schwenzer, K.\ 2010, Journal of Physics G Nuclear Physics, 37, 125202

\noindent\hangindent=0.5cm\hangafter=1 Alford, M.~G., 
Mahmoodifar, S., \& Schwenzer, K.\ 2012a, PhRvD, 85, 024007

\noindent\hangindent=0.5cm\hangafter=1 Alford, M.~G., Mahmoodifar, S., \& Schwenzer, K.\ 2012b, \prd, 85, 044051

\noindent\hangindent=0.5cm\hangafter=1 Alford, M.~G., Mahmoodifar, S., \& Schwenzer, K.\ 2012c, in AIP Conf.
Proc. 1492, QCD@WORK 2012: International Workshop on Quantum Chromodynamics: Theory and Experiment, ed. L. Angelini, G. E. Bruno, G. Chiodini et al. (Melville, NY: AIP), 257

\noindent\hangindent=0.5cm\hangafter=1 Alford, M.~G., \& Schwenzer, K.\ 2013, arXiv:1310.3524 

\noindent\hangindent=0.5cm\hangafter=1 Alpar, M.~A., Pines, D., Anderson, P.~W., \& Shaham, J.\ 1984, \apj, 276, 325

\noindent\hangindent=0.5cm\hangafter=1 Andersson, N., 
Kokkotas, K.~D., \& Stergioulas, N.\ 1999, ApJ, 516, 307 

\noindent\hangindent=0.5cm\hangafter=1 Arras, P., Flanagan, E.~E., Morsink, S.~M., et al.\ 2003, \apj, 591, 1129

\noindent\hangindent=0.5cm\hangafter=1 Arnaud, K.~A.\ 1996, in ASP Conf. Ser. 101, Astronomical Data Analysis Software and Systems V, ed. G. H. Jacoby \& J.
Barnes (San Francisco, CA: ASP), 17

\noindent\hangindent=0.5cm\hangafter=1 Bildsten, L.\ 1998, ApJL, 
501, L89 

\noindent\hangindent=0.5cm\hangafter=1 Bildsten, L., \& Ushomirsky, G.\ 2000, \apjl, 529, L33 

\noindent\hangindent=0.5cm\hangafter=1 Bogdanov, S.,
  Rybicki, G.~B., \& Grindlay, J.~E.\ 2007, \apj, 670, 668

\noindent\hangindent=0.5cm\hangafter=1 Bogdanov, S.\ 2013, ApJ, 
762, 96 

\noindent\hangindent=0.5cm\hangafter=1 Bondarescu, R., Teukolsky, S.~A., \& Wasserman, I.\ 2007, \prd, 76, 064019 

\noindent\hangindent=0.5cm\hangafter=1 Bondarescu, R., Teukolsky, S.~A., \& Wasserman, I.\ 2009, \prd, 79, 104003

\noindent\hangindent=0.5cm\hangafter=1 Brown, E.~F., \& Cumming, A.\ 2009, \apj, 698, 1020

\noindent\hangindent=0.5cm\hangafter=1 Brown, E.~F., \& Ushomirsky, G.\ 2000, \apj, 536, 915

\noindent\hangindent=0.5cm\hangafter=1 Cash, W.\ 1979, \apj, 228, 939 

\noindent\hangindent=0.5cm\hangafter=1 Chakrabarty, D., 
Morgan, E.~H., Muno, M.~P., et al.\ 2003, Natur, 424, 42 

\noindent\hangindent=0.5cm\hangafter=1 Cheng, K.~S., Chau, W.~Y., Zhang, J.~L., \& Chau, H.~F.\ 1992, \apj, 396, 135

\noindent\hangindent=0.5cm\hangafter=1 Desvignes, G.,
  Caballero, R.~N., Lentati, L., et al.\ 2016, \mnras, 458, 3341

\noindent\hangindent=0.5cm\hangafter=1 Durant, M., Kargaltsev, 
O., Pavlov, G.~G., et al.\ 2012, ApJ, 746, 6 

\noindent\hangindent=0.5cm\hangafter=1 Gonzalez, D., \& Reisenegger, A.\ 2010, \aap, 522, A16

\noindent\hangindent=0.5cm\hangafter=1 Gusakov, M.~E., Chugunov, A.~I., \& Kantor, E.~M.\ 2014, Physical Review Letters, 112, 151101

\noindent\hangindent=0.5cm\hangafter=1 Gusakov, M.~E., Chugunov, A.~I., \& Kantor, E.~M.\ 2016, arXiv:1610.06380

\noindent\hangindent=0.5cm\hangafter=1 Harding, A.~K., \& Muslimov, A.~G.\ 2002, \apj, 568, 862

\noindent\hangindent=0.5cm\hangafter=1 Haskell, B., \& Andersson, N.\ 2010, \mnras, 408, 1897

\noindent\hangindent=0.5cm\hangafter=1 Haskell, B., Andersson, N., \& Passamonti, A.\ 2009, \mnras, 397, 1464

\noindent\hangindent=0.5cm\hangafter=1 Haskell, B., Degenaar, 
N., \& Ho, W.~C.~G.\ 2012, MNRAS, 424, 93 

\noindent\hangindent=0.5cm\hangafter=1 Haskell, B., Glampedakis, K., \& Andersson, N.\ 2014, \mnras, 441, 1662

\noindent\hangindent=0.5cm\hangafter=1 Heinke, C.~O., Rybicki, G.~B., Narayan, R., \& Grindlay, J.~E.\ 2006, \apj, 644, 1090

\noindent\hangindent=0.5cm\hangafter=1 Ho, W.~C.~G., Andersson, N., 
\& Haskell, B.\ 2011, PhRvL, 107, 101101 

\noindent\hangindent=0.5cm\hangafter=1 Kantor, E.~M., Gusakov, M.~E., \& Chugunov, A.~I.\ 2016, \mnras, 455, 739

\noindent\hangindent=0.5cm\hangafter=1 Kargaltsev, O., Pavlov, G.~G., \& Romani, R.~W.\ 2004, \apj, 602, 327

\noindent\hangindent=0.5cm\hangafter=1 Larson, M.~B., \& Link, B.\ 1999, \apj, 521, 271 

\noindent\hangindent=0.5cm\hangafter=1 Levin, Y., \& Ushomirsky, G.\ 2001, \mnras, 324, 917

\noindent\hangindent=0.5cm\hangafter=1 L{\"o}hmer, O., Lewandowski, W., Wolszczan, A., \& Wielebinski, R.\ 2005, \apj, 621,
  388

\noindent\hangindent=0.5cm\hangafter=1 Lorimer, D.~R.\ 2008, Living Reviews in Relativity, 11, 8

\noindent\hangindent=0.5cm\hangafter=1 Lundgren, S.~C., Foster, R.~S., \& Camilo, F.\ 1996, in ASP Conf. Ser. 105, IAU Coll. 160, Pulsars: Problems and Progress, ed. S. Johnston, M. A. Walker, \& M. Bailes (San Francisco, CA: ASP), 497

\noindent\hangindent=0.5cm\hangafter=1 Mahmoodifar, S., \& Strohmayer, T.\ 2013, ApJ, 773, 140 

\noindent\hangindent=0.5cm\hangafter=1 Manchester, R.~N., Hobbs, G.~B., Teoh, A., \& Hobbs, M.\ 2005, \aj, 129, 1993

\noindent\hangindent=0.5cm\hangafter=1 Morrison, R., \& McCammon, D.\ 1983, \apj, 270, 119

\noindent\hangindent=0.5cm\hangafter=1 Page, D.\ 2009, Astrophysics
  and Space Science Library, 357, 247

\noindent\hangindent=0.5cm\hangafter=1 Patruno, A.\ 2010, ApJ, 722, 
909 

\noindent\hangindent=0.5cm\hangafter=1 Patruno, A., \& Watts, A.~L.\ 2012, arXiv:1206.2727 

\noindent\hangindent=0.5cm\hangafter=1 Potekhin, A.~Y., Chabrier, G., \& Yakovlev, D.~G.\ 1997, \aap, 323, 415

\noindent\hangindent=0.5cm\hangafter=1 Rangelov, B., Pavlov, G.~G., Kargaltsev, O., et al.\ 2017, \apj, 835, 264

\noindent\hangindent=0.5cm\hangafter=1 Reisenegger, A.\ 1995, \apj, 442, 749

\noindent\hangindent=0.5cm\hangafter=1 Rezzolla, L., Lamb, F.~K., Markovi{\'c}, D., \& Shapiro, S.~L.\ 2001, \prd, 64, 104014

\noindent\hangindent=0.5cm\hangafter=1 Rezzolla, L., Lamb, F.~K., \& Shapiro, S.~L.\ 2000, \apjl, 531, L139

\noindent\hangindent=0.5cm\hangafter=1 Schwenzer, K., Boztepe, T., G{\"u}ver, T., \& Vurgun, E.\ 2017, \mnras, 466, 2560

\noindent\hangindent=0.5cm\hangafter=1 Shibazaki, N., \& Lamb, F.~K.\ 1989, \apj, 346, 808

\noindent\hangindent=0.5cm\hangafter=1 White, N.~E., \& Zhang, W.\ 1997, ApJL, 490, L87 

\noindent\hangindent=0.5cm\hangafter=1 Willems, B., \&
  Kolb, U.\ 2002, \mnras, 337, 1004

\noindent\hangindent=0.5cm\hangafter=1 Yakovlev, D.~G., \& Pethick, C.~J.\ 2004, \araa, 42, 169 

\noindent\hangindent=0.5cm\hangafter=1 Yao, J.~M., Manchester, R.~N., \& Wang, N.\ 2017, \apj, 835, 29

\noindent\hangindent=0.5cm\hangafter=1 Zavlin, V.~E.\ 2006, ApJ, 638, 951 

}


\end{document}